# A Visual Analytic Environment to Co-locate Peoples' Tweets with City Factual Data

Snehal Patil* and Shah Rukh Humayoun[†]
Department of Computer Science, San Francisco State University, USA.

## ABSTRACT

Social Media platforms (e.g., Twitter, Facebook, etc.) are used heavily by public to provide news, opinions, and reactions towards events or topics. Integrating such data with the event or topic factual data could provide a more comprehensive understanding of the underlying event or topic. Targeting this, we present our visual analytics tool, called VC-FaT, that integrates peoples' tweet data regarding crimes in San Francisco city with the city factual crime data. VC-FaT provides a number of interactive visualizations using both data sources for better understanding and exploration of crime activities happened in the city during a period of five years.

## 1 INTRODUCTION

The availability of social media data, such as Twitter (currently renamed ax *X*) or Facebook data, has opened new opportunities for researchers to understand peoples' opinions and behaviors, including their perspectives on cities. Twitter's real-time updates and user-generated contents provide valuable insights towards events, news, and community experiences within a city. Integrating this data with the city factual data would open a comprehensive understanding of urban activities and issues, which could be useful for fostering a safer and more informed community.

In this work, we target at providing a visual analytics (VA) tool, called **VC-FaT** (**V**isual **C**o-locating **Fa**ctual Data with **T**weet Data), that integrates Twitter data with city factual data on crimes in a city (i.e., San Francisco) to provide insights about how people engage and discuss crimes on social media and whether there is some relationship it with the city factual crime data. The VC-FaT tool uses a number of interactive visualizations, using both data sources (i.e., the city factual crime data and the tweet data from Twitter) to visualize comprehensive overview of crime hotspots, potential danger zones, and areas of interest in the city. A visual analytics tool like VC-FaT would be useful for tourists and residents in a city to explore potentially dangerous areas and take appropriate precautions.

## 2 RELATED WORK

Researcher have developed VA tools using tweet data to help users explore events and places in the city or country. For example: Godwin et al. [2] introduced a technique for generating typographic maps by mapping geotagged tweets to neighborhood and street shapes in a city, where these maps provide geospatial visualization of tweet topics and sentiments. Qazi et al. [7] presented GeoCoV19, a large-scale multilingual Twitter dataset comprising over 524 million tweets collected during the COVID-19 pandemic, with the aim of discussing research implications of the dataset including addressing challenges such as identifying fake news and building disease forecast and surveillance models. Kozlowska and Steinnocher [3] explored the use of geotagged Twitter data to understand urban activities and define urban function in the absence of land-use information. Chae et al. [1] described an interactive visual analytic approach to extract and examine abnormal events within various social media data sources using seasonal-trend decomposition. Scholz and Jeznik [6] focused on analyzing tourist flows in Styria, Austria, using Twitter data collected from 2008 to 2018, while they used Hotspot Analysis and Kernel Density Estimation methods to investigate the spatial distribution of tourism-relevant tweets. Lu et al. [4] proposed a visual analytic framework for sentiment visualization of geo-located Twitter data in disaster scenarios. While Krstajic et al. [5] discussed the use of Twitter as a valuable source of real-time information on current events and presented an online method for detecting real-world incidents, such as natural disasters or man-made catastrophes by analyzing Twitter data. In our work, we focus on combing the tweet data with city factual data in the resulting VA environment.

## 3 THE DATASETS

We used data from two data sources, i.e., Twitter and San Francisco (SF) crime data. For San Francisco crime related tweet data, we used Twitter API[1] and collected the tweets based on crime related keywords between January 1, 2018 and December 12, 2022. The tweets were collected, 45,353 tweets, based on either having the San Francisco's districts geolocations or having the crime related keywords with mentioning San Francisco or one of its districts in the tweet. The SF crime data was collected from the official website of the SF Government[2], provided by the SF Police Department (SFPD), and covers the same period of January 1, 2018, to December 12, 2022. The dataset comprised of 602,901 crime incidents classified into various categories (e.g., arson, theft, burglary, assault, fraud, robbery, car theft, etc.). After removing the minor traffic violence records (e.g., signal crossing tickets, roadside tickets, etc.), we came up with 401,449 recorded crime incidents. The dataset contains 34 columns of information, including the neighborhood area, incident category, latitude, longitude, police district, and the time the incident was reported and occurred. In preprocessing the tweet data, we used Google Maps API[3] to map tweets to their corresponding districts in San Francisco. We used the NLTK Toolkit[4] to tokenize the keywords so to find out the crime type associated to each tweet.

## 4 VC-FAT: VISUAL CO-LOCATING FACTUAL DATA WITH TWEET DATA

We developed our VC-Fat tool as an interactive visual analytics environment to enable users exploring crime data in a city (i.e., San Francisco) with peoples' tweet data over the period of 5 years. We use a number of interactive visualizations with filtering options to explore and relate crime and tweet data targeting SF districts.

VC-FaT gives the option to view and explore data from both data sources side-by-side views or from both data sources using just one view. For example, Figure 1 shows the crimes count and tweets count using the heatmap in side-by-side view of SF all district areas.

*e-mail: snehalp31996@gmail.com
[†]e-mail: humayoun@sfsu.edu

[1]https://developer.twitter.com/en/docs/twitter-api
[2]https://data.sfgov.org/
[3]https://maps.googleapis.com/maps/api/geocode/json
[4]https://www.nltk.org/

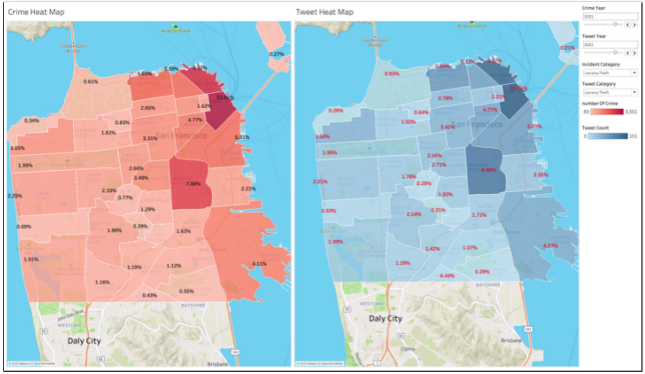

Figure 1: Left Heatmap shows San Francisco factual crime rate while right-side heatmap shows the tweet data, both between the period of January 01, 2018, to December 12, 2022.

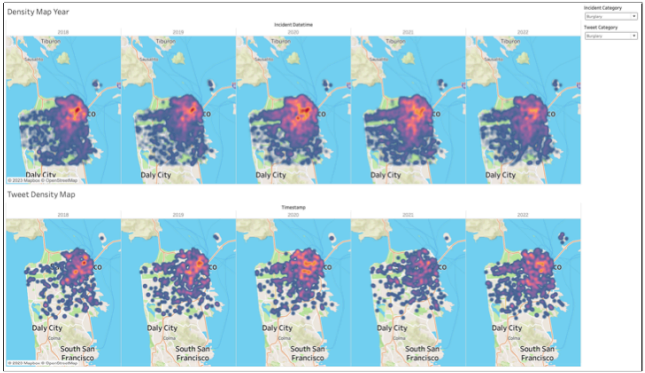

Figure 2: Top density charts show year-wise timeline of SF factual crime data while lower density chart show year-wise tweet data to each SF district.

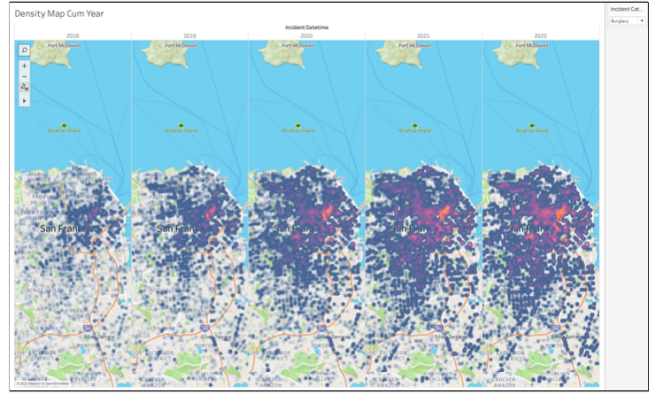

Figure 3: Accumulative evolution of crime records in SF through adding the crime counts from previous years to each next year.

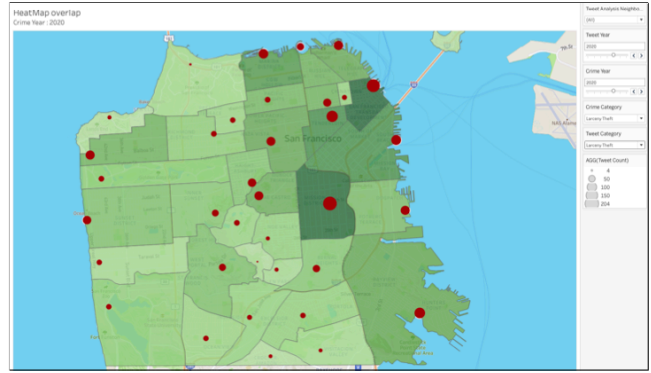

Figure 4: Heatmap showing the SF crime data while overlayed bubbles show the associated tweet data.

The left-side heat map shows the factual crime data aggregative form over 5 years and the right side shows the tweets data associated to each district. It is interesting that we see nearly the same pattern in both heatmaps towards crime effected areas. Mouse hover a particular neighborhood opens a tooltip with further details, e.g., neighborhood name, total number of crimes or tweets, etc. Users can filter the data based on one particular crime type.

VC-Fat enables users to see the timeline of both datasets with different options using density chart, e.g., Figure 2 shows the year-wise timeline of both datasets while Figure 3 shows the accumulative evolution of crime records in SF by adding the crime counts from previous years to each next year. Users can view both datasets based on days, weeks, months, or years.

VC-Fat also provides interactive visualizations to show both datasets in the same view to give users the freedom of exploring them together from different perspectives. For example, Figure 4 uses the heatmap to show the geographic distribution of crime incidents while overlayed bubbles to show the tweet data to those SF district areas. This view combines the factual crime data with public perceptions and concerns about crime in the underlying geolocation.

## 5 FUTURE WORK

In this work, we presented our VC-Fat tool to visually explore the crime activities in San Francisco districts during a period of 5 years using peoples' tweet data and the city factual crime data. In the future, we plan to expand the geographical coverage to other cities and countries as well as incorporating data from various other sources to gain a more comprehensive understanding of crime situations. We also plan to provide in-depth exploration of peoples' tweets such as sentiment analysis or emotion analysis. Furthermore, we intend to perform user studies to find out how users can understand the crime situations in city districts from the provided visualizations. Finally, we also intend to utilize such data to develop a prediction model for predicting crime activities in a city district.